# AERODYNAMIC MODELS FOR HURRICANES

# III. Modeling hurricane boundary layer

A.I. Leonov

The University of Akron, Akron, OH 44325-0301

E-mail: **leonov@uakron.edu**

## Abstract

The third paper of the series (see previous ones in Refs.[1-2]) discusses basic physical processes in the (quasi-) steady hurricane boundary layer (HBL), develops an approximate airflow model, establishes the HBL structure, and presents integral balance relations for dynamic and thermodynamic variables in HBL. Models of evaporation and condensation are developed, where the condensation is treated similarly to the slow combustion theory. A turbulent approximation for the lower sub-layer of HBL is applied to the sea-air interaction to establish the observed increase in angular momentum in the outer region of HBL. A closed set of balance relations has been obtained. Simple analytical solution of the set yields expressions for the basic dynamic variables - maximal tangential and radial wind speed components in hurricane, maximal vertical speed in eye wall, the affinity speed of hurricane travel, and the maximal temperature increase after condensation. Estimated values of the variables seem to be realistic. An attempt is also made to describe the radial distributions of wind velocity and surface pressure observed in the hurricane Frederic (1979).

## 3.1. Introduction

The HBL is located between the sea (or land) surface and the hurricane adiabatic layer. The studied below are the structure, dynamics and physical processes in the HBL of a steady (or quasi-steady) hurricane moving over open sea. These processes make overwhelming contribution in the hurricane functioning, being more complicated than those in the adiabatic layer discussed in [2]. It should be mentioned that unlike the well analyzed vertical structure of atmospheric boundary layer in planetary scale (e.g. see review in [3] Ch6), specifics of structure and processes in HBL are poorly understood.

We consider in the following the HBL having variable, radius dependent height $h_b(r)$ with $h_{max} \approx 1000-1800 m$. It is convenient to fractionate the entire HBL domain into two radial regions - the *region* 1 $(0 < r < r_e)$, and the *region* 2 $(r_e < r < r_a)$. Additionally, the region 1 is separated into the eye region $(0 < r < r_i)$ and EW region $(r_i < r < r_e)$. Hereafter we use the notation $r_e \equiv r_{e0}$. The top of the EW region 1 in HBL is the hurricane bottom spot in the upper adiabatic layer. According to employed model [1, 2] the mass, heat and momentum are supplied upwards to the hurricane EW jet only through this boundary.

In each radial region HBL can also be vertically fractionated in two sub-layers. The basic dissipative turbulent processes of heat, mass and momentum exchange between the hurricane and ocean are mostly confined in a relatively thin near-sea sub-layer with highly turbulized air whose thickness $h_s$ being $\sim 50-300 m$. The airflow structure in HBL above the turbulent sub-layer is assumed to be coherent and described by 3D aerodynamic equations. There is also a dynamic feedback from the ocean to the air that makes enormous impact on HBL dynamics.

Two basic interactions are important in the HBL. The first is the sea-air interaction. It supplies humidity due to evaporation and directly affects the HBL airflow dynamics, highly contributing to both the radial and vertical distributions of dynamic and thermodynamic variables. The second is the radial interaction of HBL with the environmental air and a possible warm air band [1]. The latter supplies the mass and heat to the EW jet in the adiabatic layer via HBL airflow. It is convenient to analyze processes in a coordinate system, moving horizontally with hurricane travel speed. In this coordinate system, the warm air band coming into the HBL creates asymmetric radial distributions of variables in the *region* 2 of the layer (see respective asymmetric model in [4]). This asymmetry will be ignored in our rough modeling.

The effects of air-sea interaction localized in the turbulent sub-layer of HPL are poorly understood. All the models so far were based on the modeling the mean wind vertical profile using logarithmic formula with the roughness parameter evaluated by scaling. The most successful approach was elaborated in paper [5].

More detailed approaches started from Phillips theory [6] of oceanic waves generated by (and moving in the same direction as) turbulent wind. More detailed and complicated, but still horizontally 1D model of the sea-air interaction was developed in publications [7, 8]. They highly extended the initial Phillips approach, involving in analysis the evolution equations for momentum and kinetic energies of turbulent pulsation and mean motion in the surface atmospheric layer. The works [7, 8] analyzed, however, only relatively low intense winds, when the oceanic waves are not broken.

It seems that the breaking oceanic waves play the dominant role in hurricane air-sea interaction, making direct dynamic contribution to both the near water turbulence and evaporation. Recent numerical studies [9] established evolutionary criteria of breakage of wind waves and were confirmed in lab experiments [10]. Nevertheless impressive results of these papers cannot be applied to hurricane dynamics where the wind-sea interactions proceed with seemingly quasi-periodic broken waves. Using a Kolmogorov-type turbulent model, recent paper [11] exposed a direct dynamic effect of ocean spray. This model predicted the turbulent drag reduction caused by the small water droplets hovering just over the wavy ocean surface. However, this model should also be further developed for possible application to hurricanes. The ocean spray can also increase by an order of magnitude the level of evaporation near the hurricane EW [12, 13].

Remaining part of the paper is organized as follows. Section 3.2 describes the aerodynamic model in the upper coherent sub-layer in HBL. This model is matched with the simplified turbulent model in the lower sub-layer in Section 3.3. Effects of evaporation and condensation are considered in Section 3.4. Section 3.5 derives integral balance relations in the HBL EW for mass, sensible and latent heat, and entropy. Section 3.6 analyzes the integral balance relations and estimates characteristic values of dynamic and thermodynamic variables. Section 3.7 attempts to describe the radial distributions of wind and surface pressure for hurricane Frederick (1979). Finally, the Section 3.8 makes a critical review the results obtained in the paper.

## 3.2. Aerodynamic models of airflow in HBL

Below aerodynamic model which intends to roughly describe airflow in HBL, is hopefully suitable for the upper, coherent sub-layer of HBL. The model employs simplified equations of aerodynamics of ideal gas (e.g. see equations (1.7-9) in Ref. [1]):

$$\partial_r(\rho r u_r) + \partial_z(\rho r u_z) = 0, \quad \partial_r(\rho r u_r M_b) + \partial_z(\rho r u_z M_b) = 0, \quad u_\varphi \approx M_b / r \qquad (3.1)$$

For simplicity the contributions of angular velocity $\Omega$ of Earth rotation in the second and third equations (3.1) are ignored. It means that the model is restricted to a radial region located not far away from the hurricane eye/EW region. In solving (3.1) for HBL the variations of density will be commonly neglected, meaning $\rho \approx \rho_a^0$.

Bearing in mind that $h_e / r_e \ll 1$ (typically 2-5%) we will use instead of vertical momentum balance (1.9) from [1] in EW region, the simplified static relation for pressure, whose integral form is:

$$p(r,z) \approx \hat{p}(r, h_b), \qquad \hat{p} = p\big|_{\rho \to \rho_a^0}. \qquad (3.2)$$

Here $\hat{p}(r, h_b)$ is the barometrically corrected radial pressure distribution at the bottom of adiabatic layer, described by formulas (2.16), (2.17) in [2], and $h_b$ generally depending on radius $r$, is the upper boundary of EW in HBL, neglecting a thin condensation layer.

Introducing then the stream function will reduce the problem to satisfying all possible continuity conditions in the upper layer of HBL. Finally, the angular momentum will be established as a linear function of the stream function [2]. Although solution of equations for ideal liquid does not satisfy the vertical boundary conditions in the whole HBL, this will be remediated by matching the aerodynamic and turbulent solutions.

### 3.2.1. Air speed distributions in the upper *region* $1\{r \leq r_e\}$ of HBL

The HBL in this region is geometrically presented as a cylinder of constant height $h_e$ with the eye $\{0 \leq r \leq r_i\}$ and EW $\{r_i \leq r \leq r_e\}$ regions (Fig.3.1). The basic assumptions for airflow in this region, similar to that in paper [2] are:

(i) $\{u_r, u_z, u_\varphi\} = \{0, 0, Ar\}$, $\{0 \leq r \leq r_i\}$ ; and (ii) $u_z = u_z(z)$ $\{r_i \leq r \leq r_e\}$. \qquad (3.3)

Evident boundary conditions provide continuity for tangential and radial wind speed components through the boundary $r = r_i$, and for all three wind speed components at the boundary $z = h_e$. We introduce non-dimensional space variables $\xi = r/r_e$, $\eta = z/h_e$, parameter $\xi_i = r_i/r_e$. The stream function $\Psi$ is commonly introduced as:

$$ru_r = -\partial_z \Psi, \quad ru_z = \partial_r \Psi. \tag{3.4}$$

The first integral $M_b = F(\Psi)$ existing for (3.1), can be well represented by the linear relation: $M_b = c_1 \Psi + c_2$, where $c_k = const$ [1,2]. The modeling condition (ii) in (3.3) and continuity of the dynamic variables at $r = r_i$ and the upper boundary $z = h_e$ provide the solution of (3.1) in the EW *region 1* $\{r_i \leq r \leq r_e, 0 \leq z \leq h_e\}$ of HBL:

$$\Psi = \frac{\xi^2 - \xi_i^2}{1 - \xi_i^2} U r_e h_e f(\eta), \quad u_r = -\frac{\xi - \xi_i^2/\xi}{1 - \xi_i^2} U f'(\eta), \quad u_z = \frac{2Uh_e/r_e}{1 - \xi_i^2} f(\eta)$$

$$M_b = \frac{\xi^2 - \xi_i^2}{1 - \xi_i^2}(1-\alpha)Mf(\eta) + \alpha M, \quad u_\varphi = \frac{M}{r_e}\left[\frac{\xi - \xi_i^2/\xi}{1 - \xi_i^2}(1-\alpha)f(\eta) + \frac{\alpha}{\xi}\right] \tag{3.5}$$

$$\xi_i \leq \xi \leq 1, \quad 0 \leq \eta \leq 1$$

Here $M = M_e$ and $\alpha \approx 0.5(r_i/r_e)^2$ have the same values as in the EW jet in adiabatic layer, $U$ is the total radial air flux through the boundary $r = r_e$ (or $\xi = 1$). This flux has two components: one induced by the radial flow in the adiabatic layer and another due to the horizontal travel of hurricane.

Function $f(\eta)$ should satisfy the evident boundary condition $f(0) = 0$. Other two boundary conditions $f(1) = 1$, and $Uf'(1) = -u_{re}$, provide the continuous transition of vertical and radial air speed components into the adiabatic layer: $\tilde{u}_z\big|_{z=h_e} = u_0$ and $\tilde{u}_r(r_e, h_e) = u_{re}$. Using these conditions and formulas in (3.5) presents the boundary values for dynamic variables at $z = h_e$ as:

$$u_0 = \frac{2Uh_e/r_e}{1-\xi_i^2}, \quad u_r\big|_{z=h_e} = u_{re}\frac{\xi - \xi_i^2/\xi}{1-\xi_i^2}, \quad M_b\big|_{z=h_e} = \frac{\xi^2-\xi_i^2}{1-\xi_i^2}(1-\alpha)M + \alpha M. \tag{3.6}$$

In the eye *region 1* ($0 < r < r_i$), the only existing wind speed component $u_\varphi$ describes due to (3.3) the quasi-solid rotational distribution,

$$u_\varphi = \alpha M r / r_i^2 \quad (0 \leq r \leq r_i).$$

Formulas (3.5) show that in the EW region, rotating air, coming from the external EW surface through the area $2\pi r_e h_e$, spreads in radial direction up to impenetrable EW inner surface $r = r_i$. This air ascends to the upper HBL level $z = h_e$, with a thin layer of condensation. It means that the rotating airflow in the EW region of HBL turns from the almost radial at the bottom to the almost vertical at the top of HBL. In fact, the first formula in (3.6) presents the integral balance relation for this airflow in the *region* 1 of HBL. At the inner boundary $r = r_i$ of HBL, formulas (3.5) predict the jump in $u_z$ describing the "frontal zone" [2]. This is a possible source of turbulization caused by the Kelvin-Helmholtz instability.

### 3.2.2 Air speed distributions in the upper *region* 2 $(r > r_e)$ of HBL

We assume the asymptotic expression $M_b \sim \sqrt{r}$. It is approximately derived below from the wind/wave interaction model in the lower turbulent sub-layer. A rough model in the *region* 2 is based on the same set of equations (3.1). Once again, the stream function $\Psi(r, z)$ is introduced according to (3.4). Then using the first integral for angular momentum $M_b = F(\Psi)$ employed in the linear form, solution (3.1) is searched as:

$$\Psi = b(r)\varphi(z), \quad M_b = a_1 \psi + a_2. \tag{3.7}$$

Here $b(r) \sim \sqrt{r}$ and $a_1, a_2$ are positive constants. Formula (3.7) shows that the HBL in the *region* 2 must be of variable height $h(r)$ to provide well documented constancy of angular momentum $M$ in adiabatic layer [2]. Since function $\Psi$ describes a streamline at $z = h(r)$, both $\Psi$ and $M_b$ are constants at $z = h(r)$. Thus the boundary conditions at the upper HBL boundary $z = h(r)$ are:

$$\Psi\big|_{z=h(r)} = \Psi_h = const, \quad M_b\big|_{z=h(r)} = a_1 \Psi_h + a_2 = M, \quad h(r_e) = h_e. \tag{3.8}$$

It is additionally assumed that in the whole HBL, including turbulent sub-layer, $\Psi \to 0$ when $z \to 0$. This condition, along with the first one in (3.7) guarantees the radial mass flux in the HBL to be constant.

Additional, "horizontal" conditions at upper boundary of HBL are:

$$\Psi(r_e, h_e) = U h_e, \quad -[\partial \Psi / \partial z]_{r_e, h_e} = u_{re}, \quad M_b(r_e, h_e) = M. \tag{3.9}$$

The stream function in the upper, aerodynamic sub-layer of HBL is modeled as:

$$\Psi(r, z) = h_e r_e U \sqrt{\xi} f(\eta), \quad U = -(u_{re} + w_*) > 0 \quad (\xi = r/r_e, \; \eta = z/h_e). \tag{3.10}$$

Parameter $w_* \approx w/\pi$ is a "pseudo-radial" contribution of hurricane translational speed $w$ in the total radial velocity. Introducing boundary conditions (3.8) in (3.10) yields:

$$f(1) = 1, \quad f'(1) = \kappa \equiv u_{re}/(u_{re} + w_*) > 0. \tag{3.11}$$

First formulas in (3.8) and in (3.9) guarantee the continuity of the mass flux at the boundary $r = r_e$ between the *regions* 1 and 2 of HBL, while the second and the third ones in (3.9) provide the continuity of radial velocity and angular momentum at $r = r_e$ in the adiabatic layer. Using (3.10), the solution of equations (3.1) in the upper sub-layer of *region* 2 of HBL is presented as:

$$\Psi(r, z) = h_e r_e U \sqrt{\xi} f(\eta), \quad u_r = -\frac{U}{\sqrt{\xi}} f'(\eta), \quad u_z = \frac{h_e}{r_e} \frac{U}{2 \xi^{3/2}} f(\eta)$$

$$M_b = M f(\eta)(1-\alpha)\sqrt{\xi} + M\alpha, \quad u_\varphi = \frac{M(1-\alpha)}{r_e \sqrt{\xi}} f(\eta) + \frac{M\alpha}{r_e \xi} \tag{3.12}$$

Comparison of (3.5) and (3.12) shows that the functions $\Psi, u_r, M_b$, and $u_\varphi$ are continuous at $\xi = 1$ (or $r = r_e$) and hold the same value of numerical factor $\alpha$. Nevertheless the vertical velocity $u_z$ has a jump at $\xi = 1$. This jump is the common feature of our aerodynamic models has been found at the boundaries of EW jet in [2].

The height $h(r)$ of HBL is established from the condition $\Psi(r, h(r)) = const$. Differentiating this condition with respect to $r$ yields the kinematical relation:

$$z = h(r): \quad u_r dh/dr = u_z. \tag{3.13}$$

The inequalities $u_r < 0$ and $u_z > 0$, which hold in HBL including its boundary $z = h(r)$, yield $h'(r) < 0$. It means that the height of HBL decreases from its maximal value at $r = r_e$ with increasing radius. The boundary conditions (3.9) at $z = h(r)$, rewritten with the aid of (3.12) yield:

$$f(\hat{h}(\xi)) = \xi^{-1/2} \quad \text{or} \quad \hat{h} = f^{-1}(\xi^{-1/2}). \tag{3.14}$$

Here $\hat{h}(\xi) = h(r)/h_e$. Since in the upper layer $f'(\eta) > 0$, the inverse function $f$ does exist and is unique. According to (3.14) $\hat{h}(\xi)$ is decreasing; the slower $f(\eta)$ increases the sharper is $\hat{h}(\xi)$ decrease. One can also establish that the external boundary of adiabatic EW jet smoothly continues downward to the HBL upper boundary $h(r)$ below the level $z = h_e$. These results show that the air trajectories in HBL, including its upper boundary $h(r)$ look like ascending centripetal spirals.

Function $f(\eta)$ characterizing vertical profiles of dynamic variables is unknown, and the aerodynamic model equations (3.1), (3.2) cannot be used to evaluate it. Its detail properties are not needed, however, for the balance relations and consequent evaluations of hurricane parameters, which is main result of the present paper. We assume below of *very small variations of $f(\eta)$, i.e. almost its constancy, except a small vicinity of $\eta = 1$*. This assumption is in fact the result of the fundamental assumption of very fast decay of radial velocity downwards, which is not justified in this paper.

Consider finally the change in the hurricane structure during hurricane traveling, when the radius of EW jet slowly changes with time, i.e. $r_e = r_e(t)$. It causes respective time dependences $M = M(t)$ and $U = U(t)$. This non-steady motion of hurricane is quasi-stationary when at any given fixed altitude $z_*$ the radial profiles of horizontal components of velocity are *self-similar,* i.e. belong to the same "master curve". This type of behavior has been observed for hurricane Frederick [14]. Using (3.12) the conditions of self-similarity in the quasi-steady motion of hurricane are written as:

$$\{M(t), u_{re}(t), w_*(t)\} \cdot \sqrt{r_e(t)} \approx \{M(0), u_{re}(0), w_*(0)\} \cdot \sqrt{r_e(0)}. \qquad (3.15)$$

This self-similarity has asymptotic character when the contribution of term $\alpha M$ in angular momentum and tangential velocity is ignored. In this case only one of relations in (3.15) (say first) is independent. If e.g. the external radius of EW $r_e$ decreases with time, the both components of horizontal wind, and the speed of hurricane travel respectively increase. Recalling assumption of the horizontal width $H$ of warm air band, $H \approx 2r_e$, we can interpret the decrease in $r_e$ as corresponding decrease in the heat supply of hurricane, whose proper maintaining causes in turn accelerated hurricane motion.

## 3.3. Modeling airflows in turbulent sub-layer of HBL

In the turbulent sub-layer, a huge air wind near the external radius $r = r_e$ maintains highest amplitude oceanic broken waves, which create a surge in EW bottom and propagate outside this region. Interaction of these waves with air changes the radial distributions of dynamical variables in the HBL. The radial wind contribution can be neglected in this sub-layer because of assumed very slow variation of $f(\eta)$. A small direct dynamic effect of vertical airflow due to evaporation at the sea surface is also negligible as compared to the values of characteristic air speed components.

As mentioned, no reliable turbulent model is currently known to describe equilibrium interaction of airflow with broken oceanic waves. Thus an empirical approach will be used below, based on the fact that at the anemometer height $z_a (\approx 10m)$ the horizontal wind speed is equal almost 75% of the air speed at the level of aircraft observation (see e.g. paper [14] and references there). This fact can happen because of direct dynamic effect of ocean spray [11]. Several results can be obtained from this observation.

Consider first the turbulent airflow in the vicinity $r = r_e$, approximated as horizontally homogeneous one. The vertical distribution of turbulent mean tangential wind $u_e$ can be described as:

$$u_e(z) = \begin{cases} A u_{*e} \ln(z/z_0), & z \geq z_0 \\ 0, & z < z_0 \end{cases}. \qquad (3.16a)$$

Here $A \approx 2.4$ is the universal value of reciprocal Karman constant, $u_{*e}$ is the dynamic velocity commonly defined via the interface shear stress $\tau_e = \rho u_{*e}^2$, and $z_0$ is the roughness factor. One can also use the common bulk relation $\tau_e = \rho C_D u_{eT}^2$ where $u_{eT}$ is the mean velocity at the height of turbulent boundary layer $h_{eT}$ and $C_D \sim 10^{-3}$ is the friction coefficient. Note that using the slight variations of vertical profile $f(\eta)$ yields the relation $u_{eT} \approx u_e$. Using both presentations of the bottom shear stress rewrites the bulk relation in another common form: $u_{*e} \approx u_e \sqrt{C_D}$. Substitution this formula into (3.16a) yields:

$$u_e(z) \approx A u_e \sqrt{C_D} \ln(z/z_0) \quad (z \geq z_0). \tag{3.16b}$$

Utilizing (3.16b) and observations [14], the values of $z_a$ and $h_{eT}$ can be found from the conditions $A\sqrt{C_D} \ln(z_a/z_0) \approx 0.75$ and $A\sqrt{C_D} \ln(h_{eT}/z_0) \approx 1$ as:

$$z_0 \approx z_a \exp\left(-\frac{0.75}{A\sqrt{C_D}}\right), \qquad h_{eT} \approx z_0 \exp\left(\frac{1}{A\sqrt{C_D}}\right). \tag{3.17}$$

Using the values $z_a = 10m$, $C_D \approx (2-4) \times 10^{-3}$ (see e.g. [15], p.71), and $A = 2.4$, the values of parameters $z_0$ and $h_{eT}$, were calculated due to (3.17) for different values $C_D$ and are shown in Table 1.

*Table 1*: Values of roughness parameter $z_0$ and height of turbulent boundary layer $h_{eT}$

| $z_a$ (m) | $C_D \times 10^3$ | $z_0$ (m) | $h_{eT}$ (m) |
|---|---|---|---|
| 10 | 1 | 0.000511 | 270 |
| 10 | 2 | 0.00923 | 103 |
| 10 | 3 | 0.0333 | 70.0 |
| 10 | 4 | 0.0714 | 51.9 |

Consider now inhomogeneous turbulent airflow in the EW bottom $r_i < r < r_e$. Using the same assumptions, one can obtain the distribution of tangential velocity similar to (3.16b) with following modification:

$$u_\varphi(z,r) \approx A u_T(\xi) \sqrt{C_D} \ln(z/z_0) \quad (z \geq z_0); \quad u_T(\xi) \approx u_\varphi(h_T, \xi). \tag{3.16c}$$

Because $h_T \approx h_e$, using the above assumptions yields,

$$u_\varphi(z_a, \xi) \approx 0.75 u_{\varphi m} \left[(1-\alpha)\frac{\xi - \xi_i^2/\xi}{1-\xi_i^2} + \frac{\alpha}{\xi}\right] \quad (\xi_i < \xi \leq 1) \tag{3.18}$$

Hereafter, $u_{\varphi m} = M/r_e$. Using the observations [14] yields $f_a = f(z_a/h_e) \approx 3/4$.

We now consider another inhomogeneous turbulent airflow outside the EW HBL region ($r > r_e$). The most intriguing dynamic effect observed in this region is the decrease in tangential velocity $u_\varphi$ with radius as $\sim 1/\sqrt{r}$, i.e. increase of angular momentum $M_b$ with radius as $\sim \sqrt{r}$. Using thermodynamic arguments, paper [16] explained observed asymptotic relation $u_\varphi \sim 1/\sqrt{r}$ assuming that the height of HBL is radius independent. This necessitated the vertical air flow through the upper boundary of HBL. In contrast to [16], the current paper explains the dependence $u_\varphi \sim 1/\sqrt{r}$ by coupling effect of propagation of oceanic waves and highly turbulized near-sea atmospheric layer.

Consider the oceanic waves propagating from the vicinity $r = r_e$ into the *region* 2. These waves, being in the equilibrium with atmospheric conditions, are propagated along the straight lines tangential to the circle $r = r_e$ with the constant phase speed $c_e$ formed at the radius $r_e$ (Fig.3.2). Therefore there is a skew interaction of the waves with the air, resulted in dominant tangential airflow in the turbulent layer. It is assumed that the mean radial turbulent wind near the sea interface is negligible. Then the tangential shear stress $\tau$ along the wave path changes from the initial value $\tau_e = \tau_{r\varphi}(r_e)$ to the value $\tau_r = \tau_{r\varphi}(r)$ at the current radius $r$, as:

$$\tau_r \approx \tau_e \sin\theta(r), \quad \sin\theta(r) = r_e/r \quad (r \geq r_e) \tag{3.19}$$

Formulas (3.19) are explained in Fig. 3.2, where the arrows at $r = r_e$ indicate the direction of propagating oceanic waves emitted from this circle. Using (3.19) and assuming that the previous results of turbulent modeling are locally applicable to the air/wind motions at any radius $r$, yields:

$$\tau_r = \rho_a^0 u_{*r}^2; \quad u_{*r} \sim c_r, \quad u_{*r} \approx u_{*e}\sqrt{r_e/r}, \quad c_r \approx c_e\sqrt{r_e/r}, \quad u_\varphi^T \sim u_\varphi \sim \sqrt{r_e/r}. \tag{3.20}$$

Here the low indexes "*e*" and "*r*" denote the values of variables at the radii $r_e$ and $r$, respectively, and $c_r$ is the local phase velocity of wave. Note that $c_r$ is not directed tangentially to the circle of radius $r$ but under angle $\hat{\theta}(r)$ defined as: $\sin\hat{\theta}(r) = \sqrt{\sin\theta(r)}$. Hereafter the upper index "*T*" denotes the values of variables in the turbulent sub-layer of HBL *region* 2. Formulas (3.20) explain the observed behavior of the tangential wind by the equilibrium-type interaction of oceanic waves with atmospheric wind, caused by the oceanic waves propagating from the surge of region 1. This friction-like interaction shows that in the turbulent sub-layer *region* 2 of HBL, the propagating oceanic waves rather *generate* air wind than *dissipate* it.

The radial component of wave action might create essential radial airflow in the turbulent layer. However, this possible airflow would be directed in positive radial direction, just opposite the centripetal radial flow in the upper aerodynamic sub-layer. Unstable shear band created by these two types of radial flow produces the Kelvin-Helmholtz instability with very high level turbulent pulsations, intermixing the sub-layer, and creating seemingly very low (if any) mean radial flow.

Using the same assumptions the mean turbulent speed distribution can be approximately described by (3.16c), however changing (3.18) for

$$u_\varphi(z_a, \xi) \approx 0.75 u_{\varphi m}[(1-\alpha)/\sqrt{\xi} + \alpha/\xi] \quad (\xi \geq 1). \tag{3.21}$$

Formulas (3.18) and (3.21) will be used in estimations of evaporation.

## 3.4. Evaporation, balance of latent heat, and condensation jump

### 3.4.1. Effects of evaporation

Over calm oceanic waters, the vertical air flux (per unit mass of vapor) caused by moisture evaporation can be approximated as [15] (p.67):

$$u_v(r) = C_E |\underline{V}_a|, \quad C_E = C_e(q_{sea} - q). \tag{3.22}$$

Here $\underline{V}_a$ is the wind speed at the anemometer level $z_a$ (~10m), $C_E \approx 2 \times 10^{-3}$ the exchange coefficient, and $q_{sea} - q_{air}$ ($\approx 2.4 g/kg = 2.4 \times 10^{-3}$) is the difference between specific humidity of air and that near the sea surface. This difference characterizes the weight concentration of vapor. Thus over calm oceanic waters, $C_E \approx 5 \times 10^{-6}$.

The ocean spray ejected over the oceanic water by wave whitecaps, can increase by an order of magnitude the value of $u_m(r)$ at the hurricane EW [12]. The above modeling of turbulent sub-layer cannot, however, evaluate the horizontal region where the wave whitecaps exist. Hence in the following evaluations of evaporation region we use the observed facts [15] that evaporation effectively ceases at the radius $r_{ex} \approx 150 - 350 km$. Neglecting the radial component of turbulent air flow and taking into account that $u_\varphi$ at the hurricane bottom is always directed cyclonically, simplifies formula (3.22) as: $u_v(r) \approx C_E u_\varphi(z_a, r)$.

Using (3.18) and (3.21), the total mass flux of evaporation is calculated as:

$$Q_v = 2\pi\rho_v C_E \int_{r_i}^{r_{ex}} r u_\varphi(z_a, r) dr \approx 1.5\pi\rho_v u_{\varphi m} r_e^2 C_E \left[ \alpha(\xi_{ex} - \xi_i) + \frac{(1-\alpha)}{3}\left(\frac{1+\xi_i - 2\xi_i^2}{1+\xi_i} + 2(\xi_{ex}^{3/2} - 1)\right) \right]$$
(3.23)

Here $\rho_v$ is the vapor density and $\xi_{ex} = r_{ex}/r_e$.

### 3.4.2. Integral balance of latent heat

According to the present model the vapor flux comes to the upper EW region and is solely responsible for the vapor condensation there.

The latent heat due to evaporation is delivered from the oceanic surface to the adiabatic EW jet entrance. This heat is released in condensation under cooling and pressure drop in the ascending air in the EW jet. In calculation of the latent heat we use the approximate relation (3.23) for the total vapor flux $Q_v$, neglecting seemingly small amount of latent heat release due to the condensation of small spray droplets near the ocean water. The integral balance of latent heat is:

$$\pi(r_e^2 - r_i^2)\rho_0 u_0 \Delta l_e \approx Q_v \Delta l \quad (l = L_v m).$$
(3.24)

Here $L_v$ is the specific latent heat of vaporization, $m$ is humidity, $\Delta l$ and $\Delta l_e$ are the values of latent heat excess over the environmental air at the sea surface and that coming to the EW jet. The left-hand side of (3.24) is the excess of latent heat coming upwards to the adiabatic EW jet being supplied from evaporation at hurricane sea surface.

### 3.4.3. Transition of HBL to the adiabatic layer: Condensation

The side heat supply to hurricane can not in general provide enough temperature increase to overcome the stability threshold for existing (quasi) adiabatic jet above the HBL (see [2]). So a rough scheme below shows a way for establishing the value $T_0$ of temperature coming into the bottom of adiabatic EW jet. The physics underlying the scheme is that the latent heat accumulated in vaporized air and transferred to the upper level of *region*1 of HBL, suddenly releases with supply of additional sensible heat. It seems that without this type of heat transfer the normal functioning of hurricane is impossible. This scheme means that the height $h_e$ of HBL is restricted by the condensation level $h_c$, which could be roughly evaluated under simplified condition that condensation starts upon reaching the saturation point ([3], p.175):

$$h_c \approx 0.12(T_- - T_{D0}) \quad (km) \tag{3.25}$$

Here $T_- - T_{D0}$ is the dew point temperature depression at the sea surface. Usually, $T_- - T_{D0} \approx (10-15)^0 C$, so $h_c \approx 1.2 - 1.8 km$ (see [3], Fig.6.2, p.166).

It is assumed that the condensation happens in a relatively thin vertical layer whose height is less than hundred meters, where the over-saturated vapor comes into the upper layer of HBL. This thin layer could be considered as a weak condensation jump (e.g. see [17]), neglecting the thickness of the layer. The basic equations for the condensation jump, which include the conservation of vertical fluxes of mass, momentum and energy, have been established in [17]. For weak jump these equations reminding the combustion equations, are reduced to the continuity of mass flux, pressure and enthalpy on the jump interface, averaged over EW radius:

$$\rho_- u_- = \rho_0 u_0, \quad p_- = p_0, \quad E_- = c_p T_- + <m> L = c_p T_0 = E_0 \tag{3.26}$$

It can be established that the difference between $u_0$ and $u_-$ is negligible being equal about several cm/sec units. So in the following one can approximate $u_0 \approx u_-$ and use for pressures $p_0$ and $p_-$ the equation for ideal gas. Those simplifications are the same as in slow combustion theory [17]. Therefore the only evaluation for $<m>$ is needed to use equation (3.26). It readily comes for the oceanic vapor mass balance as:

$$Q_v = \pi \rho_v (r_e^2 - r_i^2) u_- <m> . \tag{3.27}$$

Here the left-hand side is the mass vapor flux coming from ocean due to evaporation and the right-hand side presents the flux of over-saturated vapor at the height $h_c$ of condensation surface.

Assuming that the all vapor from the ocean surface under hurricane is completely condensed in the condensation layer, using (3.23), (3.27) and approximation $u_0 \approx u_-$, yields:

$$<m> \approx \Upsilon C_E u_{\varphi m} / u_0 \qquad (3.28)$$

Here the non-dimensional geometrical factor $\Upsilon$ is:

$$\Upsilon = \frac{1.5}{1-\xi_i^2}\left[\alpha(\xi_{ex}-\xi_i) + \frac{(1-\alpha)}{3}\left(\frac{1+\xi_i-2\xi_i^2}{1+\xi_i} + 2(\xi_{ex}^{3/2}-1)\right)\right] \qquad (3.29)$$

With the aid of (3.28) and (3.29) the last relation in (3.26) is rewritten as:

$$T_0 = T_- + <m> L/c_p . \qquad (3.30)$$

This temperature increase due to condensation seems to be the most essential part of heat supply in the upper, adiabatic hurricane EW jet.

In the region $z > h_c$ the two-phased air–water mixture contains the small droplets formed by condensation of oceanic vapor. Centrifugal forces move the droplets out of EW in the external region with formation of specific rain bands. We do not consider further the formation of the rain bands which needs an additional analysis (see discussion and review in Ref. [15], Sect. 2.3.6).

## 3. 5. Integral balances in the region 1 of HBL

Apart from the thermodynamic parameters $T_a^0$, $p_a^0$, which characterize the standard non-perturbed adiabatic (ambient) atmosphere outside the hurricane, other parameters at the bottom of adiabatic layer, such as the average temperature $T_0$ at the bottom spot, vertical velocity $u_0$, the velocity $w$ of the horizontal motion of hurricane, and the value of angular momentum $M$ in adiabatic layer are still unknown. Thus given the geometrical structure of hurricane, these unknown parameters are evaluated from integral balances formulated below.

In the following we use the horizontal orthogonal axes $x$, $y$; the axis $x$ being coincided with the axis of the warm air band. Then the horizontal ($\|$) and normal ($\perp$) projections of the sailing and affinity components of horizontal wind are:

$$w_{\|} = w_{s\|} - w_a, \quad w_{\perp} = w_{s\perp}. \tag{3.31}$$

### 3.5.1. Mass balance in HBL

According to the present model, the air mass coming to the EW jet at the bottom of adiabatic layer from HBL is provided by two sources: (i) the entrainment into the HBL of radial airflow from the bottom of adiabatic layer (or the upper layer of HBL), induced by the inclination of external boundary of EW jet there, and (ii) the involvement of fresh air by the horizontal traveling of hurricane. Then neglecting density variations, the mass balance is presented as:

$$\pi(r_e^2 - r_i^2)u_0 \approx 2\pi r_e h_e |u_{re}(r_e)| + H h_e |w|, \quad |w| = \sqrt{(w_{s\|} - w_a)^2 + w_{s\perp}^2} \tag{3.32}$$

Here we used the first, mass balance relation (3.26) in the condensation sub-layer, neglecting the difference between $u_-$ and $u_0$, $H \approx 2r_e$ is the width of warm air band, and the vapor flux $Q_v$ (3.23) being neglected in (3.32). The left-hand side of (3.32) is the mass flux of air leaving the HBL, i.e. entering the bottom of hurricane adiabatic EW jet. The first term in the right-hand side of (3.32) describes the rate of mass supplies by induced radial flow in boundary layer, and the third one from the warm air band coming to the region 2 of HBL caused by both sailing wind and affinity speed. According to this model, all of these air parcels are transported into the EW jet from the *region* 2 through the side boundary of HBL at $r = r_e$.

Recall that $u_{re} = u_0 r'_{re}$ and that due to (2.31) and (2.32) [2],

$$r'_{re} \approx -k(\tau_0 - \tau_c)/\tau_2, \quad k = g/(7RT_a^0 \sqrt{q_0 + 1}), \quad \tau_c \equiv \tau_1 \approx 2.5\mu_0(1 + /q_0)\varphi_1(q_0)$$

$$\tau_2 = \mu_0(1 + 1/q_0)\varphi_2(q_0), \quad \tau_0 = 3.5(T_0 - T_a^0)/T_a^0, \quad \mu_0 = u_{\varphi m}^2/(2RT_a^0). \tag{3.33}$$

Here $r'_{re}$ is the initial slope of adiabatic EW jet, $q_0 = (r_e/r_i)^2 - 1$, and the functions $\varphi_1(q_0)$ and $\varphi_2(q_0)$ are given by formulas (A1) and (A2) in Appendix 2 paper [2]. Remarkably, the initial inclination of EW jet affects the increase in temperature $T_0$ after condensation. It helps to overcome the critical temperature $T_c$ of the stability threshold at the bottom of adiabatic EW jet.

### 3.5.2. Balance of sensible heat in the EW region of HBL

We consider the sea/air interface in complete temperature balance, where the sea temperature $T_s$ is equal to that of near-water air, i.e. the ambient temperature $T_a^0$. Because the direct heat exchange between ocean and HBL is proportional to $T_s - T_a^0$, it doesn't contribute to the sensible heat balance. It is in accord with the Chapter 3.9 of text [8] for stationary (developed) wave air interaction, and in contrast with the assumption of some numerical models (e.g. see [18]). The assumption $T_s \approx T_a^0$ is, however, violated when the dissipation heat is taken into account. It is shown at the end of Section 5.3 that the dissipative heat increases near-the-sea-level temperature by $\approx 0.1\,^0C$.

Neglecting molecular and turbulent heat transfer in the *region* 1 of HBL leaves advection of temperature as the only remaining heat exchange process,

$$\partial_r (r u_r \tau) + \partial_z (r u_z \tau) = 0, \quad \tau = 3.5(T - T_a^0)/T_a^0.$$

Here $\tau$ is the non-dimensional temperature introduced in [2]. Neglecting the thermal effect of dissipative heating and using the aerodynamic formulas for velocity, the general solution of this equation is obtained as $\tau = G(\Psi) + const$. Here $\Psi$ is the stream function in *region* 1 of HBL, defined in (3.5). Using as before the linear relation $\tau \sim \hat{\Psi}$ yields:

$$\tau = \tau_+ \frac{w_{*a}}{U} \frac{\xi^2 - \xi_i^2}{1 - \xi_i^2} f(\eta). \tag{3.34}$$

Here $\tau_+$ is corresponding non-dimensional temperature at the warm air band, $w_{a*}$ is the pseudo-radial horizontal speed of hurricane due to the affinity, $U$ is the integral speed of radial motion in HBL. Formula (3.34) shows that there is no heat supply to the hurricane eye region $r < r_i$ because $\tau\big|_{\xi=\xi_i} = 0$. Due to the boundary condition $f(0) = 0$ formula (3.34) is also in accord with our assumption $T_{oc} = T_a^0$.

The radial average of (3.34) should be equal to value $\tau_-$, corresponding to the temperature $T_-$ at $\eta = 1-0$, i.e. just before the condensation temperature jump. Note that the notation $\tau_0$ used in [2] corresponds to the temperature $T_0$ at $\eta = 1+0$ just after the condensation jump. Because of (3.30) $\Delta T = T_0 - T_- > 0$.

Calculating the radial average of (3.33) at $\eta = 1-0$ yields:

$$\tau_- = \tau_+ w_a / U .\qquad(3.35)$$

Formula (3.35) can also be obtained from the horizontal balance of thermal flux where the direct heat exchange between ocean and air and effects of dissipative heating are neglected. Here the advective thermal flux is the only significant heat source that comes from the warm air band and supplies the heat to EW below the condensation layer. Then the balance of *sensible heat* in stationary hurricane takes the form:

$$\pi(r_e^2 - r_i^2)c_{p0}u_0\tau_- \approx Hh_e c_{p+} w_a \cdot \tau_+ \quad \left[\tau_- = 3.5(T_-/T_a^0 - 1),\ \tau_+ = 3.5(T_+/T_a^0 - 1)\right].\qquad(3.36)$$

Here $c_{p0} \approx c_{p+}$ are the heat capacities of air just before the condensation jump and at the warm air band, respectively; $T_-, T_a^0$ and $T_+$ are respective temperatures at $z = h_c - 0$ in hurricane EW, in the hurricane environment, and at the warm air band. The left-hand side of (3.35) describes the heat entering the hurricane condensation layer, and the right-hand side there describes the air heat supplied by the warm air band, which comes to the hurricane only due to the affinity. Evidently (3.36) is equivalent to (3.35).

### 3.5.3. Entropy balance and dissipation in the region 1 of HBL

The entropy balance with dissipative friction was first used by K. Emmanuel for evaluating maximum air speed in hurricane (e.g. see [19, 20] and references there). He roughly analyzed the closed paths of air parcels from the near sea water to the troposphere as a Carnot cycle with dissipation caused by friction of wind at the ocean surface. A lot of effort was made fitting the calculations with results of observations, paying the most attention to the ocean spray effect on the sea-atmosphere heat exchange. This work still continues.

Unlike the modeling [19, 20], we use the entropy balance to establish a relation between maximum wind speed $u_{\varphi m}$ and vertical initial speed $u_0$ of adiabatic EW jet. This sub-section analyses the entropy balance with frictional dissipation only in the region 1, since at the bottom of region 2 the kinetic energy of wind is seemingly not dissipated but increases due to the wind-wave interaction. This generation airflow by sea waves is perhaps the main reason for the small wind energy lost in HBL.

We use in the following the standard approach of non-equilibrium thermodynamic [21], with the well-known general entropy balance relation formulated as:

$$\partial_t(\rho\Delta S)+\underline{\nabla}\cdot(\rho\Delta S\underline{v})=-\underline{\nabla}\cdot(\underline{J}_q/T)+P_s$$
$$TP_s=-T^{-1}\underline{J}_q\cdot\underline{\nabla}T+\underline{\underline{\sigma}}:\underline{\underline{e}}-\rho dF/dt\big|_T \tag{3.37}$$

Here $S$ and $F$ are the entropy and Helmholtz free energy densities (per mass unit), $\underline{J}_q$, $\underline{\underline{\sigma}}$ and $\underline{\underline{e}}$ are the molecular thermal flux, stress and strain rate tensors, respectively, $\underline{v}$ is the velocity vector, and $P_s$ is the *entropy production*, which according to the Second Law of thermodynamics is positive for all non-equilibrium processes.

Neglecting the density variations and molecular thermal flux, and introducing the dissipation localized at the dissipative (turbulent) layer as the "friction", we can rewrite the entropy balance (3.37) for steady hurricane as:

$$\partial_r(ru_r\Delta S)+\partial_z(ru_z\Delta S)=P_s\equiv D_{loc}/T\,; \quad D_{loc}\approx C_D r u_\varphi^3 \tag{3.38}$$

Here $D_{loc}$ is the local dissipation presented by the "bulk friction" formula. We further use the formula (1.2) from [1] for entropy, normalized on ambient conditions:

$$\Delta S=c_p\ln(T/T_a^0)+L_v m/T-R\ln(p/p_a^0).$$

Here $R$ is the humid air gas constant, $p$ and $p_a^0$ are hurricane and ambient pressures, $L_v$ and $m$ are specific latent heat and specific humidity, respectively. Bearing in mind small relative deviations of temperature and pressure in HBL from the ambient values, we can represent the above relation in a simplified form:

$$\Delta S\approx c_p\tau/3.5+L_v m/T_a^0+R(p_a^0-p)/p_a^0. \tag{3.39}$$

Integrating stationary equation of entropy balance (3.37) over the domain $\{z_a<z<h_e,\ r_i<r<r_e\}$ of the EW region of HBL yields:

$$\rho u_0\int_{r_i}^{r_e}[\Delta S(h_e,r)-f_a\Delta S(z_a,r)]rdr-\rho r_e h_e\int_{\eta_a}^{1}|u_r(z,r_e)|\Delta S(z,r_e)d\eta\approx\frac{C_D}{T_a^0}\int_{r_i}^{r_e}ru_\varphi^3(h_e,r)dr \tag{3.40}$$

Using the fact that the sensitive and latent heat contributions in the entropy have been balanced (see integral balances above), the only remaining contribution in the entropy $\Delta S$ is the pressure contribution $\Delta S_p\approx R(p_a^0-p)/p_a^0$. Because of (3.2) this quantity is altitude independent. In (3.40) the air speed components $u_r(z,r_e)$ and $u_\varphi(h_e,r)$ are defined in (3.5), and $f(\eta_a)=f_a\approx 3/4$ as based on the approximate results of Section 3.3.

We now use formula (2.16) [2] neglecting $u_{re}$ contribution. It presents the radial distribution of quantity $p_a^0 \Delta S_p / R \approx p_a^0 - p$ as:

$$p_a^0 - p = \frac{\rho u_{\varphi m}^2}{2}\left(1 + \frac{s_e Z(s)}{(\Delta s)^2}\right) \qquad (3.41)$$

Here $s = r^2$, $\Delta s = s_e - s_i$, $s_i = r_i^2$, $s_e = r_e^2$, and

$$Z(s) = (1-\alpha)^2(s_e - s) + (s_i - \alpha s_e)^2 \frac{s_e - s}{s s_e} - 2(1-\alpha)(s_i - \alpha s_e)\ln\frac{s_e}{s} \qquad (3.42)$$

In our rough calculations below we neglected the adiabatic pressure variations and density changes with altitude. It particularly means that $\rho \approx \rho_a^0$. It is seen from (3.41) and (3.42) that $p_a^0 - p = \rho u_{\varphi m}^2 \Delta s / 2$ at $r = r_e$. Calculations of (3.40) in Appendix 1, show

$$u_0 = \lambda u_{\varphi m} \quad (\lambda = \lambda_0 C_D). \qquad (3.43)$$

Here parameter $\lambda_0$ depends only on the ratio $r_i / r_e$. General calculations of $\lambda_0$ are presented by awkward formula (A1.2) shown in Appendix.

Note that the right-hand side of (3.41) multiplied by $2\pi T_a^0$ represents the total mechanical dissipation $D$ at the bottom of *region* 1. Using (A1.1) its expression can be written as:

$$D = 2\pi s_e C_D u_{\varphi m}^3 \int_{\xi_i}^{1}\left(\alpha + (1-\alpha)\frac{\xi^2 - \xi_i^2}{1-\xi_i^2}\right)^3 \frac{d\xi}{\xi^2}. \qquad (3.44)$$

In particular case, $r_i / r_e = \xi_i = 1/2$ ($\alpha = 1/8$), easy calculation using formulas (A1.2) and (A1.3) from Appendix 1 show:

$$\lambda_0 \approx 11.7, \quad D \approx 0.306\pi s_e C_D u_{\varphi m}^3. \qquad (3.45)$$

### 3. 5.4. Affinity speed of hurricane

Consider the situation when the sailing wind contribution to the hurricane travel is negligible. Then the horizontal hurricane travel entirely depends on affinity. A simple scaling description of affine motion of hurricane is based on the assumption of proportionality of travel speed $w_a$ to the affinity force $\tau_+$ with a coefficient proportional to a characteristic speed, say $u_{\varphi m}$:

$$w_a = \nu u_{\varphi m} \tau_+ . \tag{3.46}$$

Here $\nu$ is a positive constant which may depend on the geometrical parameters of hurricane. Assuming parcels of warm air band being effectively mixed by dominant tangential air speed resulted in formula (A2.1) from Appendix 2. Then using definition $\tau_+ = 3.5 \Delta T_+ / T_a^0$ formula (A2.1) yields (3.46) where $\nu = 2\pi/3.5 \approx 1.8$.

## 3. 6. Analysis of balance relations in HBL and calculations

### 3.6.1. General relations and particular cases

There are five unknown characteristic parameters in HBL: the temperatures $T_-$ and $T_0$ just before and after condensation jump, maximal tangential speed $u_{m\varphi}$, the initial velocity $u_0$ of adiabatic EW jet, and contribution of affinity speed $w_a$ in the horizontal speed of hurricane. The components of sailing wind $w_s$ and the horizontal temperature supply $T_+$ are considered as known. The balance relations include: (i) condition at the condensation jump (3.30) (with additional relations (3.28), (3.29)), (ii) mass balance (3.32), (iii) sensible heat balances (3.36), (iv) entropy balance resulted in formula (3.43), and (v) the definition of affinity speed (3.46). These balance relations present the closed set of five algebraic equations for determining the above five parameters.

Recall that the non-dimensional temperatures, $\tau_+$, $\tau_-$ and $\tau_0$ are defined as:

$$\tau_+ = 3.5(T_+ - T_a^0)/T_a^0, \quad \tau_- = 3.5(T_- - T_a^0)/T_a^0, \quad \tau_0 = 3.5(T_0 - T_a^0)/T_a^0 .$$

It is convenient to use furthermore the non-dimensional characteristic wind components, scaled with the adiabatic speed $\sqrt{RT_a^0}$, as:

$$\{u_0, u_{\varphi m}, w_{s\|}, w_{s\perp}, w_a\} = \{u, v, \hat{w}_{s\|}, \hat{w}_{s\perp}, \hat{w}_a\}\sqrt{RT_a^0} . \tag{3.47}$$

Then the balance of temperatures (3.30) on condensation jump takes the form:

$$\tau_0 = \tau_- + (u/v)X , \quad \left(X = 3.5 \Upsilon C_E L/(c_p T_a^0)\right) . \tag{3.48}$$

In our rough calculations we do not distinguish in the following the values of densities and heat capacity in (3.32) and (3.36). Along with (3.48), using the above relations results in the three equations for three non-dimensional parameters $\tau_-, u$ and $v$ as:

$$\lambda v[1-(\gamma_1\tau_- +\gamma_2/\lambda)/v^2 +\gamma_3] = \varepsilon|\hat{w}|/\pi, \quad \tau_- = \upsilon\tau_+^2, \quad u = \lambda v. \tag{3.49}$$

Here the second equation has been obtained upon substitution (3.46) into equation (3.36) neglecting the differences between the heat capacities, $|\hat{w}| = \sqrt{(\hat{w}_{s\|}+\hat{w}_a)^2 + \hat{w}_{s\perp}^2}$, and the non-dimensional constants $\gamma_n$, $k$, $\varepsilon$, and $\upsilon$ are presented as:

$$\gamma_1 = \frac{2ks}{(1+q_0^{-1})\varphi_2(q_0)}, \quad \gamma_2 = \gamma_1 X, \quad \gamma_3 = \frac{2.5ks\varphi_1(q_0)}{\varphi_2(q_0)}, \quad \varepsilon = \frac{2h_e/r_e}{1-\xi_i^2}, \quad k = \frac{gr_i^2}{7RT_a^0 r_e}, \quad \upsilon = \frac{v\varepsilon}{\lambda\pi} \tag{3.50}$$

Here $q_0 = (r_e/r_i)^2 - 1$, and $\varphi_1(q_0)$, $\varphi_2(q_0)$ are functions tabulated in [2].

Equations (3.48) and (3.49) are closed set. Using the second equation in (3.48) and relations for $\tau_0$ and $w_a$ yields:

$$\tau_0 = \upsilon\tau_+^2 + X/\lambda, \quad \hat{w}_a = \nu v \tau_+. \tag{3.51}$$

The first term in the first formula (3.51) presents contribution of external horizontal heat source from the warm air band, and the second from the condensation.

Substituting $\hat{w}_a$ in the right-hand side of the first equation in (3.49) results in an awkward algebraic relation between the non-dimensional maximal tangential wind speed $v$, and given values of horizontal temperature $\tau_+$ and sailing wind components $\hat{w}_{s\|}$ and $\hat{w}_{s\perp}$. For illustrating purpose, only two limiting cases of this equation are considered below.

(i) *External sensible heat supply is negligible,* $\tau_+ = 0$. In this case the hurricane moves only under action of given sailing wind $\underline{w}_s$ and the affinity speed $w_a$ is negligible. Then $\tau_- = 0$, $w_a = 0$, $|\hat{w}| = \hat{w}_s$, and the use of formula $u = \lambda v$, reduces the first equation in (4.48) to the following:

$$v[1-(\gamma_2/\lambda)/v^2 +\gamma_3] = \varepsilon w_s/(\pi\lambda)$$

After solving this equation all unknown variables are presented in dimensional form as:

$$u_{\varphi m} = \frac{\varepsilon w_s + \sqrt{(\varepsilon w_s)^2 + 4RT_a^0 \pi^2 \lambda \gamma_2 (1+\gamma_3)}}{2\pi\lambda(1+\gamma_3)}; \quad u_0 = \lambda u_{\varphi m}; \quad |u_{re}| = \frac{u_0}{\varepsilon} - \frac{w_s}{\pi}; \quad \tau_0 = X/\lambda \quad (3.52)$$

In the limit case $w_s = 0$ when the horizontal motion is absent, the hurricane still exists, and the particular case of solution (3.52) is:

$$u_{\varphi m}^0 = u_{\varphi m}\big|_{w_s=0} = \sqrt{\frac{RT_a^0 \gamma_2}{\lambda(1+\gamma_3)}}; \quad u_0 = \lambda u_{\varphi m}^0; \quad |u_{re}| = \frac{u_0}{\varepsilon}; \quad \tau_0 = \frac{X}{\lambda}. \quad (3.53)$$

(ii) *Sailing wind is negligible,* $w_s = 0$. In this case when the hurricane moves with affinity speed $w_a$, the first equation in (3.49) takes the form:

$$1 + \gamma_3 - (\gamma_1 \tau_+^2 + \gamma_2/\lambda)/v^2 = \upsilon\tau_+.$$

After solving this equation all unknown variables are presented in dimensional form as:

$$u_{\varphi m}^2 = RT_a^0 \frac{\gamma_1 \upsilon \tau_+^2 + \gamma_2/\lambda}{1+\gamma_3 - \upsilon\tau_+}; \quad u_0 = \lambda u_{\varphi m}; \quad w_a = v u_{\varphi m} \tau_+; \quad |u_{re}| = u_0 \frac{1-\upsilon\tau_+}{\varepsilon}; \quad \tau_0 = \upsilon\tau_+^2 + \frac{X}{\lambda} \quad (3.54)$$

The limit case $\tau_+ \to 0$ ($w_a = 0$) once again is described by (3.53).

In analyzing obtained solutions, consider first the limit solution (3.53). The last formula shows that heat supply $\tau_0$ to the adiabatic EW jet is entirely produced by the condensation jump. The first formula in (3.52) shows that the condensation heat, proportional to $\tau_0$ (the numerator in the radical there), is balanced by the friction (denominator in the radical there). We now analyze the dependence of maximal tangential wind speed $u_{\varphi m}^0$ on the external radius $r_e$ of EW. Here we roughly assume that due to similarity of radial hurricane structure, the height of condensation and parameter $\gamma_3$ are independent of $r_e$. Then formulas (3.53) show that

$$u_{\varphi m}^0 \sim \sqrt{r_e^0/r_e}, \quad u_0 \sim \sqrt{r_e^0/r_e}, \quad |u_{re}| \sim \sqrt{r_e/r_e^0}. \quad (3.55)$$

Here $r_e^0 = 40 km$ is accepted value of $r_e$ for the "standard" hurricane. If $u_{\varphi m}^0$ has dominant contribution in $u_{\varphi m}$ under action of sailing wind and/or side heating, formulas (3.55) are in agreement with the quasi-static similarity conditions (3.15) in traveling hurricane [14].

Consider now the qualitative effects of sailing wind $w_s$ when the sensible side thermal effects are absent. Formulas (3.52) show that $u_{\varphi m}$ and $u_0$ are monotonically increasing functions of $w_s$, whereas $|u_{re}|$ might decrease with $w_s$ growing. Once again, the heat supply $\tau_0$ to the adiabatic EW jet is entirely produced by the condensation jump. In spite of dependence of variables $u_{\varphi m}$, $u_0$ and $|u_{re}|$ on $r_e$ as described by (3.55), their realistic values are very close to those calculated at very small non-dimensional values of $w_s$ (see Section 3.6.3).

Consider finally the effects of side heating $\tau_+$ with no effect of sailing wind. Formulas (3.54) show that $u_{\varphi m}$, $u_0$, $w_a$, and $\tau_0$ increase with $\tau_+$ growing, while $|u_{re}|$ as a function of $\tau_+$ might be generally non-monotonous. When the value $\tau_+$ is small enough, the dynamic variables depend on $r_e$ as shown by the formulas (3.55), where $w_a \sim \sqrt{r_e^0 / r_e}$. At higher values of $\tau_+$, the solutions depend on $r_e$ in more complicated manner.

### 3.6.2. Numerical illustrations

We remind that the unknown non-dimensional variables are the hurricane basic velocities, $v$ and $u$, and the unknown relative temperature $\tau_-$ before the condensation jump; the given (external) variables are the velocity of sailing wind $\underline{w}_s$ and the relative temperature $\tau_+$ of the warm air band. Along with these variables there are parameters $\gamma_k, \lambda$, mostly depending on the hurricane geometry, as well as the parameter $\nu$. As soon as the unknown variables are determined from solving set of equations (3.48) and (3.49), two other basic variables, the radial velocity $u_{re}$ and the relative temperature $\tau_0$ after condensation jump could easily be found too.

**(i)** *Determining parameters*

Geometrical parameters

We consider the "standard" hurricane with basic geometrical parameters: $r_i = 20km$, $r_e = 40km$, $h_e = 1.5km$, and the effective radius of evaporation $r_{ex} = 200 - 300km$. Other non-dimensional geometrical parameters calculated from the basic ones are: $q_0 = 3$, $\alpha = 1/8$, $\varphi_1(q_0) \approx 1.25$, $\varphi_2(q_0) \approx 0.033$, $\kappa \approx 0.16$, $\xi_i = 0.5$, $\varepsilon = 0.1$, and $\xi_{ex} = 5$.

Physical parameters

The value of environmental temperature is assumed as $T_a^0 = 300^0 K$ ($27^0 C$), with the velocity scaling factor $\sqrt{RT_a^0} = 295 m/s$. The values of heat capacity at constant pressure $c_p$ and latent heat of vaporization $L$ are taken from the text [22] as: $c_p = 10^3 JK^{-1}/kg$, and $L = 2.3 \times 10^6 J/kg$. The values $C_D$ and $C_E$ characterizing friction and evaporation, respectively, are: $C_D \approx 3.5 \times 10^{-3}$ and $C_E \approx (1-5) \times 10^{-5}$. Here we use high enough value for $C_D$ shown in the text [15] (p.71) characterizing strong Hurricane Inez 1966. Additionally, the value of non-dimensional parameter $\nu = 1.8$, characterizing the affinity component $w_a$ of hurricane travel speed is accepted from the Sub-Section 3.5.4, and Appendix 2.

Calculated parameters

Parameters $\gamma_1$ and $\gamma_3$ depending only on geometry are defined in (3.54) and calculated as: $\gamma_1 = 0.727$, $\gamma_3 = 1.515$. To calculate the value $\gamma_2$ represented the effect of condensation heat, we first use (3.29) to find $\Upsilon \approx 13.4$. Then using $C_E \approx (1-5) \times 10^{-5}$ and calculating due to (3.48) parameter X yields: $\gamma_2 \approx (2.6-13) \times 10^{-3}$. Using above value $C_D \approx 3.5 \times 10^{-3}$, the value of parameter $\lambda \approx 0.04$ was calculated due to the first formula in (3.45). Finally, parameter $\upsilon$ was calculated due to the last formula in (3.50) as $\upsilon \approx 1.43$.

**(ii) *Results of calculations***

We demonstrate the calculations beginning with the particular case (3.53) when the hurricane does not travel. Depending on the values of evaporation parameters $C_E$ and $\xi_{ex}$ calculation of $u_{\varphi m}^0$ yields: $u_{\varphi m}^0 \approx (43.5 - 97.6) m/s$. To eliminate uncertainty in the choice of $C_E$ we tune up the value of parameter $\gamma_2$ to obtain $u_{\varphi m}^0 \approx 50 m/s$ for the standard hurricane. This value is $\gamma_2 = 2.93 \times 10^{-3}$. Thus for $\xi_{ex} = 5$, $C_E = 1.32 \times 10^{-5}$. This value $\gamma_2$ is furthermore fixed in all below calculations with $X = \gamma_2/\gamma_1 \approx 4 \times 10^{-3}$. The values of parameters established in this sub-section for accepted hurricane geometry are summarized in Table 2. Except adjusted parameter $\gamma_2$ which reflects the effect of evaporation, all other parameters have been closely evaluated.

Table 2: *Non-dimensional parameters in calculations of standard hurricane*

| Parameter | $\lambda$ | $\nu$ | $\upsilon$ | $\gamma_1$ | $\gamma_2$ | $\gamma_3$ | $\varepsilon$ |
|---|---|---|---|---|---|---|---|
| Value | 0.04 | 1.8 | 1.43 | 0.727 | 0.00293 | 1.515 | 0.1 |

Consider first the case (i) when hurricane travels with the speed of the sailing wind $w_s$, without side supply of sensible heat. In this case $w_a = 0$ and $\tau_+ = \tau_- = 0$. The initial temperature increase $\Delta T_0$ in the adiabatic EW jet is caused only by condensation. The stability condition $\tau_0 > \tau_c$ for adiabatic EW jet is satisfied here, because the value $\tau_c \approx 0.0598$ calculated using (5.33), is less than $\tau_0 \approx 0.1$ shown in Table 3. Calculations of (3.52) with the sailing speed equal to 5, 10 and 15 $m/s$ show that within 0.1% precision, the tangential air speed remains the same (i.e. $u_{\varphi m} \approx 50 m/s$) as in case of not traveling hurricane. Then with the same precision the initial vertical speed in the adiabatic EW jet is also not affected by the hurricane travel under sailing wind. Because of the mass balance equation (the fourth formula in (3.52)), the value of maximal radial speed monotonically decreases with increasing sailing wind. It means that the fresh air coming to hurricane EW due to its traveling with sailing wind diminishes the air attraction caused by the inclination of EW boundary. Calculated values of basic variables are shown in Table 3.

Table 3: *Results of calculations of basic variables in hurricane travel under given values of sailing wind* $w_s$ ($\Delta T_+ = 0$)

| $w_s$, m/s | $u_{\varphi m}$, m/s | $u_0$, m/s | $|u_{re}|$, m/s | $\tau_-$ | $\Delta T_-^0 C$ | $\tau_0$ | $\Delta T_0^0 C$ |
|---|---|---|---|---|---|---|---|
| 0 | 50.0 | 2.00 | 20.0 | 0 | 0 | 0.100 | 8.57 |
| 5 | 50.0 | 2.00 | 18.4 | 0 | 0 | 0.100 | 8.57 |
| 10 | 50.0 | 2.00 | 16.8 | 0 | 0 | 0.100 | 8.57 |
| 15 | 50.0 | 2.00 | 15.2 | 0 | 0 | 0.100 | 8.57 |

Consider now the case (ii) when the hurricane travels without sailing wind under given values of relative side temperatures $\Delta T_+ = T_+ - T_a^0$ equal to 3, 6 and $10^0 C$. Corresponding basic variables, calculated due to formulas (3.54) are shown in Table 4.

Table 4: *Results of calculations of basic variables in affine travel of hurricane under given values of $\Delta T_+^0 C$ ($w_s = 0$)*

| $\Delta T_+$ $^0C$ | $\tau_+$ | $u_{\varphi m}$ m/s | $w_a$ m/s | $u_0$ m/s | $\|u_{re}\|$ m/s | $\tau_-$ | $\Delta T_-$ $^0C$ | $\tau_0$ | $\Delta T_0$ $^0C$ |
|---|---|---|---|---|---|---|---|---|---|
| 0 | 0 | 50.0 | 0 | 2.00 | 20.0 | 0 | 0 | 0.100 | 8.57 |
| 3 | 0.035 | 50.7 | 2.6 | 2.03 | 19.5 | 0.0014 | 0.12 | 0.1014 | 8.69 |
| 6 | 0.070 | 52.1 | 5.25 | 2.08 | 19.1 | 0.0056 | 0.48 | 0.1056 | 9.05 |
| 10 | 0.117 | 55.0 | 9.3 | 2.20 | 19.05 | 0.0157 | 1.345 | 0.1157 | 9.92 |

Comparing the results of calculations shown in Tables 3 and 4, one can notice that in the case (i) (Table 3) the only variable sensitive to the sailing wind speed is the radial velocity $|u_{re}|$. Other key variables remain almost the same as in case of not traveling hurricane. On the contrary, except $|u_{re}|$, the values of key hurricane variables grow noticeably with the increase in the sensible side heat supply (see Table 4). As shown in Tables 3 and 4, in both cases the values of radial velocity $|u_{re}|$ steadily decreases with the increase of hurricane travel velocity $w_s$ or $w_a$. It means that due to the mass balance, the increasing rate of air parcels entering the adiabatic EW jet from the HBL create the lower value of initial tangent $|r'_e|$ of the hurricane EW jet.

We finally evaluate the contribution of dissipative heat $D$ in the heat balance. Using the second formula in (3.45), the dissipation per unit surface area of EW is presented as $D/(0.75 u_0 \pi r_e^2) = 0.408 C_D u_{\varphi m}^2$. Equalizing this formula to the convective term, $c_p \Delta T_d$, yields the effective increase in temperature due to dissipation as:

$$\Delta T_d \approx 0.408 C_D u_{\varphi m}^2 /(\lambda c_p). \tag{3.56}$$

Substituting parameters in (3.56) with values $u_{\varphi m} = (50 - 55) m/s$ from Table 4 yields: $\Delta T_d \approx (0.089 - 0.118)^0 C$. These values almost do not affect the values of $u_{\varphi m}$ and $T_0$ and make noticeable contribution in $T_-$ ($\ll T_0$) only when values of $\Delta T_+$ are very small.

## 3.7. Calculations of radial profiles of wind speed and surface pressure in the hurricane Frederic (1979)

There are not so much observations reported in the literature, which provide simultaneous data for radial profiles of both the wind and surface pressure. One of them has been reported in paper [14] for hurricane Frederic that approached the Alabama-Mississippi Gulf Coast in September 1979. The data for the wind speed were obtained at the elevation 500m by aircraft and the surface pressure distributions were measured by buoys. Two of these coupled radial distribution data for consecutive two days of observation were presented in Figure 4 in paper [14], one for September 11 and another for September 12. In both cases the hurricane traveled over open sea. Some specifics of hurricane Frederic structure reported in [14] should be kept in mind when considering the comparison of our calculations with the observations in [14]. First, the hurricane Frederic travels horizontally in NNW with the speed ~ 5m/s, which created asymmetric distribution of wind and pressure. As reported in paper [14], "The wind field… was remarkable constant outside the radial distance of 40km from the storm center. The radius of maximum winds ($R_{max}$) and the maximum 30s average flight level wind ($V_{max}$) varied depending on quadrant penetrated and time. The mean $R_{max}$ decreased from ~ 33 to 27 km; the central pressure dropped from 960 to 948 $mb$ and the $V_{max}$ increased from 48 to 58 m/s from late (23:00 GMT) 11 September to early (07:00 GMT) 12 September flight, respectively. These changes are illustrated in (our!) Fig.3.3, which details profiles of surface pressure and wind speed on the north side of the storm at the 500m altitude during 11 September and 12 September flights. Outside 40 km, the radial surface pressure gradients and the flight-level wind speeds are essentially equal for both flights, while inside 40 km, the surface pressure gradient and wind speeds are stronger for the 12 September flight. Several passes in other quadrants indicated the same tendency".

We compare below our calculations for radial distributions of wind speed and surface pressure in the boundary layer with the data presented in Fig.4 of paper [14], and copied in this paper as Fig.3.3. One can see from this figure that the data for the maximum wind speeds are considerably higher than the reported average ones, and their locations are also different from averaged values. However, the minimum central pressures are almost the same as averaged values. The hurricane centers in the both data presented are unknown.

Ignoring possible contribution of radial wind speed, we use for comparison the expression $u_\varphi$ in (3.5) for the radial distribution of tangential wind speed, assuming that the vertical profile $f(\eta)$ in $u_\varphi$ is flat, so the value of angular momentum at the flight altitude is close to that in adiabatic layer. We calculated the radial pressure distribution using formulas (2.16) from paper [2]:

$$\frac{p}{p_a^0} \approx \begin{cases} 1 - v^2 \left(1 + \alpha^2 s_e (s_i - s)/s_i^2 + s_e Z(s_i)/(\Delta s)^2\right)/2, & (0 \leq s \leq s_i) \\ 1 - v^2 \left(1 + s_e Z(s)/(\Delta s)^2\right)/2, & (s_i \leq s \leq s_e) \\ 1 - v^2 s_e /(2s), & (s \geq s_e) \end{cases} \quad (3.57)$$

Here $v = u_{\varphi m}/\sqrt{RT_a^0}$, and

$$Z(s) = (1-\alpha)^2 (s_e - s) + (s_i - \alpha s_e)^2 \frac{s_e - s}{s s_e} - 2(1-\alpha)(s_i - \alpha s_e) \ln \frac{s_e}{s} \quad (3.58)$$

In (3.57) and (3.58), $s = r^2$, $s_i = r_i^2$, $s_e = r_e^2$, $\Delta s = r_e^2 - r_i^2$; $p_a^0$ and $T_a^0$ are the ambient surface pressure and temperature. The values of parameters in the calculations are:

$$\alpha = 1/8, \quad r_i = r_e/2 \quad (q_0 = 3), \quad p_a^0 = 1,000 mb, \quad RT_a^0 = 87,000 (m/s)^2. \quad (3.59)$$

In this modeling the maximal tangential speed of hurricane $u_{\varphi m}$ and the radius $r_e$ of eye jet are considered as fitting parameters, since in the original paper [14] they could not be reported precisely because of pulsations of wind speed.

The comparisons of calculations with the raw data [14] are presented in Figures 3.4.a,b and 3.5.a,b for the observations on 11 and 12 September, respectively. Here the figures 3.4.a and 3.5.a show the calculated and measured tangential velocity distributions, whereas 3.4.b and 3.5.b the pressure distributions related to 11 and 12 September, respectively. The calculations are presented by solid lines imposed on the fluctuating data copied from the data [14], with the fitted parameters of hurricane indicated in the figure captions. These parameters for data in Fig.3.4 were chosen as $u_{\varphi\max} = 62 m/s$ and $r_e = 36 km$. whereas for Fig. 3.5 they are: $u_{\varphi\max} = 72 m/s$ and $r_e = 25 km$. Neglecting possible differences between these parameters measured in the boundary layer as compared with their values at the bottom of EW jet of hurricane, we calculate the angular momentum $M_{11} \approx 2.23 \times 10^6 m^2/s$ and $M_{12} \approx 1.80 \times 10^6 m^2/s$ for both days indicated in indices. Another parameter $\mathrm{M} = u_{\varphi\max}\sqrt{r_e}$ characterizing the quasi-static evolution of radial distribution of $u_\varphi$ is considerably closer for both the observations: $\mathrm{M}_{11} \approx 372 (m/s) km^{1/2}$, $\mathrm{M}_{12} \approx 360 (m/s) km^{1/2}$. Therefore these two measured wind profiles close to each other when $r > 40 km$. Recall that contribution of radial velocity in those calculations were ignored. When this contribution in the radial distribution of the horizontal wind seems to be insignificant, ignoring it in the pressure distribution results in slightly higher pressure profiles, shifted towards the small radius values. Additionally, the maximum of tangential speed in upper adiabatic layer of hurricane might be slightly higher than accepted using the assumption of slight change of vertical profile. Therefore the comparison between the measured and computed pressure profiles can be considered as fair enough.

### 3.8. Conclusions

The present third paper of the series develops axisymmetric aerodynamic models which illustrate the stationary 3D structure of hurricane boundary layer (HBL). These models establish the distributions of dynamic variables and temperature depending on radial and vertical coordinates. The main obstacle here is poorly understood processes in turbulent sub layer and consequently, vague values of friction and evaporation bulk coefficients.

The modeling is based on the simplifying assumption that the HBL and the upper adiabatic layer exchange their mass and heat only through the upper surface of eye wall. In this modeling, the whole HBL is fractioned in the radial direction in two regions, region 1 which includes the cylindrical domain of eye and (a hollow cylindrical domain of) EW sub-regions ($r < r_e$), and the region 2 outside it, i.e. $r > r_e$ (Fig.3.1). The EW region of HBL is restricted above by the level of condensation, which is treated in spirit of slow combustion theory as a thin layer with condensation jump. Additionally each region is separated in vertical direction into turbulent lower sub-layer and upper aerodynamic layer with mostly coherent airflows.

The description of airflow in the coherent upper sub-layer is based on simplified aerodynamic equations, involving the continuity and angular momentum balance equations. In spirit of boundary layer theory, the vertical pressure distribution across the HBL is neglected. As soon as the stream function is determined, the angular momentum is found as a linear function of the stream function, using the first integral existing in these types of airflow.

The description of the turbulent airflow is based on observations that at the anemometer level (~10m), the wind speed is about ¾ of that at the aircraft (~500m) level, and matching the turbulent and aerodynamics descriptions of the airflows.

In the eye region the solid like rotation of air satisfies the aerodynamic equations. In the EW of HBL the distributions in aerodynamic layer are obtained using the assumption of well mixed, radial independent vertical velocity, with radial velocity non-penetrable into the eye sub-region. This assumption yields a description of stream function and respective solution of angular momentum balance, which corresponds to the tangential air wind. Using advection as the only heat transfer mechanism, additional expression for temperature distribution in the region 1 is obtained. Involved in this description is also bottom friction concentrated in the turbulent sub-layer.

In the region 2 of HBL, the angular momentum radial grow as $\sim \sqrt{r}$, found in observations, was derived by easy extension of theory of coupled sea waves – air wind interaction [9] (Ch.3.5) in the developed (equilibrium) state. Using this fact and continuity conditions at the interface $r = r_e$ for stream functions and angular momentums, allows describe all the dynamic variables. Continuity of stream function or angular momentum at the upper boundary $h(r)$ of HBL with adiabatic layer, where the angular momentum is constant, allows establishing the form of this boundary depending on the local upper distribution of vertical profile. It is shown that the upper HBL boundary $h(r)$ decreasing with $r$ growing, and is a smooth continuation of the external boundary of EW jet in adiabatic layer.

The model assumes no heat exchange at the interface between ocean surface and hurricane air, meaning that both temperatures coincide at the interface. It is in accord with the Chapter 3.9 of text [9] for stationary (developed) wave/air interaction and in contrast to the assumption of some numerical models (e.g. see Ref.[15]).

Although very small direct contribution of evaporation in the velocity field was neglected, the evaporation plays overwhelming role in releasing latent heat and delivering respective temperature increase to the adiabatic upper EW jet. It was shown that this heat stabilizes the EW jet and creates initial curvature in its boundary, which induces significant radial flow from the hurricane environment. Another possible source of the heat supply for stable functioning of hurricane comes from the warm air band in which direction the hurricane moves with the "affinity" speed.

The integral balance relations have been formulated for region 1 of hurricane for transfer of mass, sensible and latent heat. Using these balances, additional integral entropy balance with bottom friction was reduced to the balance of pressure flux. Calculations established proportionality between the maximal tangential and vertical velocities in EW of HBL. A phenomenological relation was also proposed which models the affinity velocity component of hurricane travel in the direction of the warm air band. Simple analytical solution of the integral balance relations allowed evaluating (under given temperature of warm air band and/or sailing wind velocity) the basic dynamic variables - maximal tangential and radial velocities in hurricane, maximal vertical speed in eye wall, the speed of hurricane travel and the maximum temperature after condensation jump at the boundary with adiabatic layer. These evaluations for a "standard" hurricane showed realistic features.

Additionally, the raw data for radial distributions of wind velocity and surface pressure observed in the hurricane Frederic (1979) were semi-quantitatively simulated using the present modeling.

**Appendix 1:** *Calculations of entropy balance*

In calculations of (3.39) we will take into account that $\Delta S_p = R(p_a^0 - p)/p_a^0$ where $p$ is altitude independent, $\rho = \rho_a^0$, $\rho_a^0 R T_a^0 = p_a^0$, and $Z(s_e) = 0$ for $Z(s)$ defined in (3.42). Then substituting in (3.41) $\Delta S_p$ from (3.42), as well as $u_r$ and $u_\varphi$ from (3.5), yields:

$$u_0(u_{\varphi m}^2/4)(1-f_a)\left(\Delta s + \int_{s_i}^{s_e} \frac{s_e Z(s)}{(\Delta s)^2} ds\right) - r_e h_e u_0 \Delta s (u_{\varphi m}^2/2)|U|(1-f_a)$$
$$= C_D u_{\varphi m}^3 s_e \int_{\xi_i}^{1} \left(\alpha + (1-\alpha)(\xi^2 - \xi_i^2)/(1-\xi_i^2)\right)^3 \frac{d\xi}{\xi^2} \quad (A1.1)$$

Equation (A1.1) immediately results in formula (3.43): $u_0 = \lambda u_{\varphi m}$. Using now the mass balance relation (3.32) (or the first formula in (3.6)) yields: $u_0 \Delta s = 2 h_e r_e U$. It means that the out-of-integral terms in the left-hand side of (A1.1) are cancelled. Thus parameter $\lambda_0$ in (3.43) has the general expression:

$$\lambda_0 = \frac{4(\Delta s)^2}{1-f_a} \int_{\xi_i}^{1} \left(\alpha + (1-\alpha)\frac{\xi^2 - \xi_i^2}{1-\xi_i^2}\right)^3 \frac{d\xi}{\xi^2} \times \left(\int_{\xi_i}^{1} Z(s) ds\right)^{-1}. \quad (A1.2)$$

In particular case $r_i/r_e = \xi_i = 1/2$ ($\alpha = 1/8$),

$$\int_{\xi_i}^{1} \left(\alpha + (1-\alpha)\frac{\xi^2 - \xi_i^2}{1-\xi_i^2}\right)^3 \frac{d\xi}{\xi^2} \approx 0.112, \quad \int_{\xi_i}^{1} Z(s) ds \approx 0.153(\Delta s)^2. \quad (A1.3)$$

The evaluations of integrals in (A.1.3) are used in formulas (3.45) of the main text.

**Appendix 2**: *On optimal orthogonal mixing of passive admixtures*

Consider homogeneous orthogonal mixing of a passive admixture of concentration $x$. The stream delivering admixture has velocity $u_1$ and admixture flux $q_1 = u_1 x$. The orthogonal stream has the take up mixing velocity $u_2$, the flux $q_2 = u_2 x$, and mixing dissipation $r_m = q_2 x = u_2 x^2$. The mixing efficiency is defined as: $e_m = q_1 - r_m = x(u_1 - u_2 x)$. The optimal problem is formulated as following: given the mixing velocity $u_2$, find delivery speed $u_1$ providing a concentration $x_o$ maximizing the mixing efficiency. Maximizing $e_m$ with respect of $x$ yields: $x_o = u_1/(2u_2)$; $\max e_m = u_1^2/(4u_2)$. If $x_o$ is known, the optimum delivery speed is: $u_1 = 2xu_2$. In applying this formula to the problem of hurricane affinity travel, one needs to make substitutions: $x = \Delta T_+ / T_a^0$, $u_2 = u_{\varphi m}$, and $u_1 = w_{*a} = w_a / \pi$. It yields:

$$w_a = 2\pi u_{\varphi m} \Delta T_+ / T_a^0 \ . \tag{A2.1}$$

This formula is used for evaluation of parameter $\nu$ in Section 3.5.4.

## Acknowledgment

The author expresses a deep gratitude to Dr. A. Benilov (Stevens Institute of Technology, NJ) for many useful suggestions and discussions. Many thanks are also given to my former Ph.D. Student A. Gagov for calculations and graphics.

**Figure Captions**

Fig. 3.1. Cross-sectional sketch of hurricane

Fig. 3.2. Sketch of radius dependent change in direction of surface shear stress along the path of propagating ocean wave.

Fig. 3.3. Radial profiles of wind speed and surface pressure on the north side of Frederic at the 500 m level for the 11 September (11 I) and 12 September (12 H1) flights (reprinted from [14], Fig.4).

Fig. 3.4. Comparison of adiabatic calculations and measurements [14] for radial distributions of tangential wind and surface pressure for hurricane Frederic for measurements on 09/11/1979: (a) tangential wind and (b) surface pressure; maximal wind speed $u_{\varphi \max} = 62 m/s$, radius of eye jet $r_e = 36 km$.. Red and blue solid lines are measurements and calculations, respectively.

Fig. 3.5. Comparison of boundary layer calculations and measurements [14] for radial distributions of tangential wind and surface pressure for hurricane Frederic for measurements on 09/12/1979: (a) tangential wind and (b) surface pressure; 09/12/1979: maximal wind speed, $u_{\varphi \max} = 72 m/s$, radius of eye jet $r_e = 25 km$. Red and blue solid lines are measurements and calculations, respectively.

**Figures**

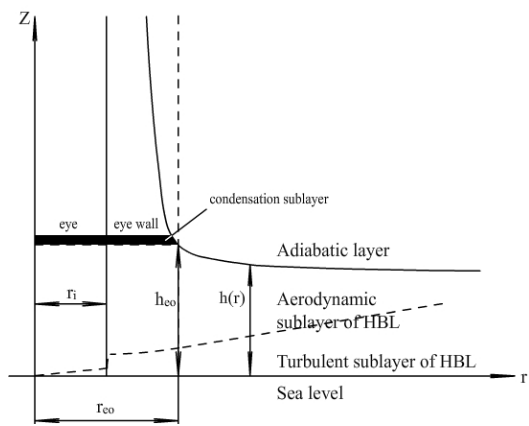
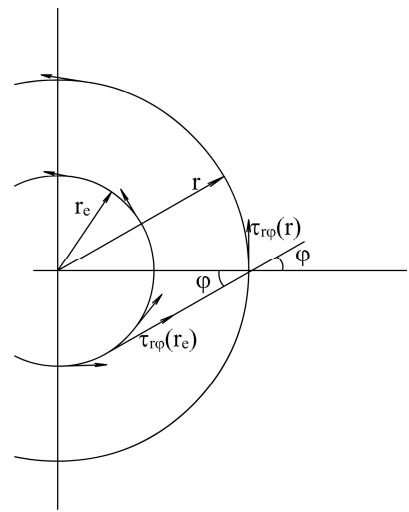

Fig.3.1            Fig.3.2

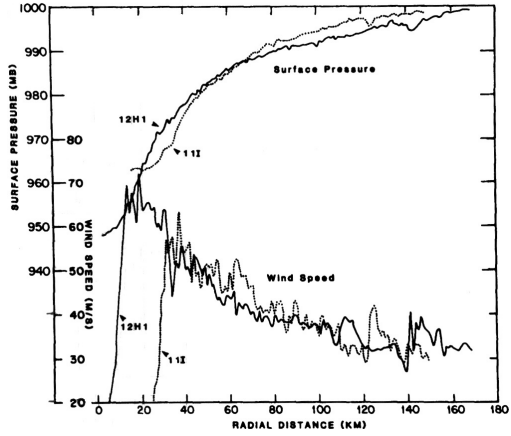

Fig.3.3

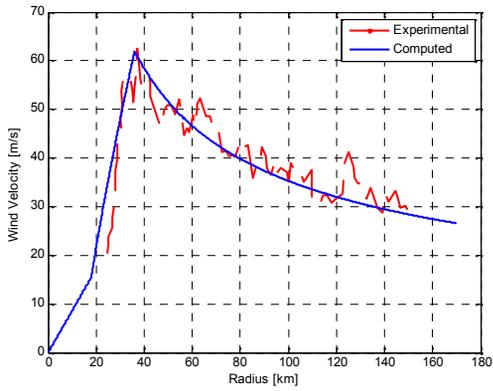

Fig.3.4a

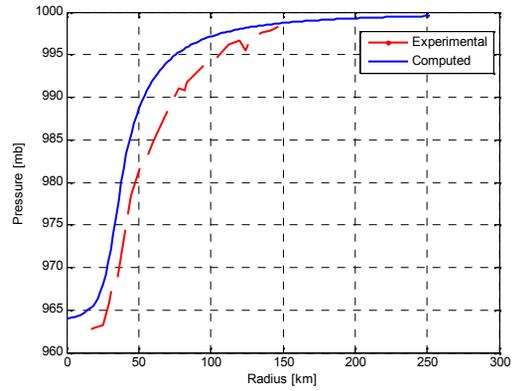

Fig.3.4b

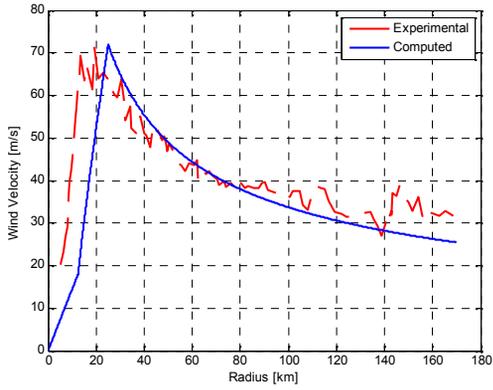

Fig.3.5a

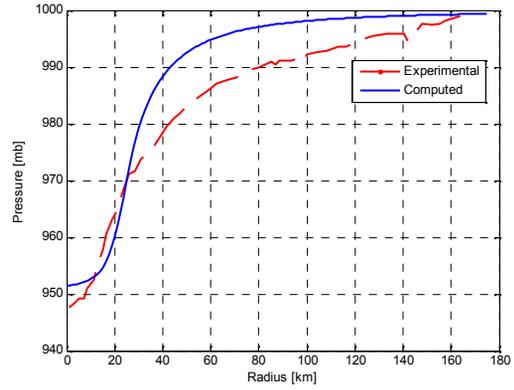

Fig.3.5b
38